\documentclass[smus]{snowhack}
\usepackage{graphics,epsf}
\begin{document}

\title{
                         Structure Function Subgroup Summary
   \thanks{
   Work supported in part by NSF and DoE.
   }
}

\author{
M.    Albrow$^{\ref{FNAL}}$,
E.L.  Berger$^{\ref{ANL}}$,
T.    Bolton$^{\ref{KSU}}$,
A.    Caldwell$^{\ref{Columbia}}$,
A.    El-Khadra$^{\ref{UIUC}}$,
B.    Ermolaev$^{\ref{Ioffe}}$,
J.W.  Gary$^{\ref{River}}$,
D.    Harris$^{\ref{Roch},}$\thanks{Subgroup Conveners},\\
E.    Hughes$^{\ref{CalTech}}$,
J.    Huston$^{\ref{MSU}}$,
D.    Jansen$^{\ref{MPG}}$,
Y.-K. Kim$^{\ref{FNAL}}$,
D.    Krakauer$^{\ref{ANL}}$,
S.    Kuhlmann$^{\ref{ANL}}$,
H.    Lai$^{\ref{MSU}}$,
D.    Naples$^{\ref{KSU}}$,\\
J.    Qui$^{\ref{ISU}}$,
F.    Olness$^{\ref{SMU},\dagger}$,
D.    Reeder$^{\ref{Wisc}}$,
M.H.  Reno$^{\ref{UIowa}}$,
S.    Ritz$^{\ref{Columbia},\dagger}$,
R.J.  Scalise$^{\ref{SMU}}$,
B.    Schumm$^{\ref{SLAC}}$,
P.    Spentzouris$^{\ref{Columbia}}$,\\
C.    Taylor$^{\ref{CWRU}}$,
W.-K. Tung$^{\ref{MSU}}$,
J.    Wang$^{\ref{SMU}}$,
X.    Wang$^{\ref{MSU}}$,
B.    Ward$^{\ref{TENN}}$,
H.    Weerts$^{\ref{MSU}}$,
A.    White$^{\ref{ANL}}$,
J.    Whitmore$^{\ref{PSU}}$,
W.    Yu$^{\ref{SMU}}$
}

\date{
\newcounter{inst}
\begin{list}{$^{\arabic{inst}}$}{\usecounter{inst}}
\centering
\renewcommand{\itemsep}{-0.15in}
\em
\item Fermi National Accelerator Laboratory,
      Batavia, IL 60510\label{FNAL}\\
\item Argonne National Laboratory, Argonne, IL 60439-4815\label{ANL}\\
\item Kansas State University, Physics Department, 116 Cardwell Hall,
      Manhattan, KS 66506\label{KSU}\\
\item Columbia University, Department of Physics,
      New York, NY 10027\label{Columbia}\\
\item University of Illinois at Urbana-Champaign, Loomis Laboratory of Physics,
      1110 West Green Street, Urbana IL 61801-3080\label{UIUC}\\
\item A.F.Ioffe Physico-Technical Institute, St.Petersburg,
      194021 Russia\label{Ioffe}\\
\item University of California at Riverside, Physics Department,
      Riverside, CA 92521\label{River}\\
\item University of Rochester, Department of Physics and Astronomy,
      River Campus, B \& L Bldg., Rochester, NY 14627\label{Roch}\\
\item California Institute of Technology, HEP - 452-48,
      Pasadena, CA 91125\label{CalTech}\\
\item Michigan State University, Department of Physics and Astronomy, 
      East Lansing, Michigan 48824-1116\label{MSU}\\
\item Max-Planck Institut fuer Kernhysik, Saupfercheckweg 1,
      69117 Heidelberg, Germany\label{MPG}\\
\item Iowa State University, Department of Physics,
      Ames, Iowa 50011-3160\label{ISU}\\
\item Southern Methodist University, Department of Physics, 
      Dallas, TX 75275-0175\label{SMU}\\
\item University of Wisconsin, Department of Physics,
      Madison, WI 53706\label{Wisc}\\
\item The University of Iowa, Department of Physics and Astronomy,
      Iowa City, Iowa 52242\label{UIowa}\\
\item Stanford Linear Accelerator Center, P.O. Box 4349,
      Stanford, CA 94309\label{SLAC}\\
\item Case-Western Reserve University, Department of Physics,
      Cleveland, OH 44106\label{CWRU}\\
\item University of Tennessee, Department of Physics,
      Knoxville, TN 37996\label{TENN}\\
\item The Pennsylvania State University, Department of Physics, 
      University Park, PA 16802-6300\label{PSU}
\end{list}
}

\maketitle


\thispagestyle{plain}\pagestyle{plain}

\begin{abstract} 
We summarize the studies and discussions of the Structure Function subgroup
of the QCD working group of the Snowmass 1996 Workshop: 
{\em New Directions for High Energy Physics}.
\end{abstract}

\newcommand{\nub}{\overline{\nu}}
\newcommand{\ubar}{\overline{u}}        
\newcommand{\dbar}{\overline{d}}
\newcommand{\alfs}{\mbox{$\alpha_s$}}
\newcommand{\asop}{\mbox{$\frac{\alpha_s}{\pi}$}}
\newcommand{\qsq}{\mbox{$Q^2$}}
\newcommand{\qnsq}{\mbox{$Q_0^2$}}
\newcommand{\mztwo}{\mbox{$M_Z^2$}}
\newcommand{\lmsb}{\mbox{$\Lambda_{\overline{MS}}$}}
\newcommand{\lmqcd}{\mbox{$\Lambda_{QCD}$}}
\newcommand{\st}{\scriptstyle}
\newcommand{\sst}{\scriptscriptstyle}
\newcommand{\mco}{\multicolumn}
\newcommand{\ra}{\rightarrow}
\newcommand{\vp}{{\bf p}}
\newcommand{\rmt}{\rm\textstyle}
\newcommand{\nunuc}{\mbox{$\nu$N}}
\def\NCA{\em Nuovo Cimento}
\def\NIM{\em Nucl. Instrum. Methods}
\def\NIMA{{\em Nucl. Instrum. Methods} A}
\def\NPB{{\em Nucl. Phys.} B}
\def\PLB{{\em Phys. Lett.}  B}
\def\PRL{{\em Phys. Rev. Lett.}}
\def\PRD{{\em Phys. Rev.} D}
\def\ZPC{{\em Z. Phys.} C}

\section{INTRODUCTION} 

Our knowledge of the structure functions of hadrons, and the parton density 
functions (PDFs) derived from them, has improved
over time, due both to the steadily increasing quantity and precision
of a wide variety of measurements, and a more sophisticated theoretical
understanding of QCD.  Structure functions, and PDFs, play a dual role:
they are a necessary input to predictions for high momentum transfer 
processes involving hadrons, and they contain important information 
themselves about the underlying physics of hadrons.  Their study is an 
essential element for future progress in the understanding of fundamental 
particles and interactions.

Because of the ubiquitousness of structure functions, the activities of the
subgroup had significant and 
productive overlap with several other subgroups, and were focused in a 
number of different directions.  This summary roughly follows
these directions.  We start with the precision of our knowledge of the PDFs.  
There was an attempt to define a `Snowmass convention' on PDF errors,
 reviewing the experimental and theoretical input to the extraction of the PDFs
and an appraisal of what is left to do.  Next, we explore the important 
connection between the strong coupling constant, \alfs, and the structure 
functions.  One of the important inputs provided by the structure functions 
is in the precise extraction of electroweak parameters at hadron colliders.  
The systematic uncertainties in the 
structure functions may be the limiting factor in the determination of 
electroweak parameters, and this is discussed in the subsequent section.  
There is then a review of some relevant aspects of heavy quark
hadroproduction.  Finally, as a summary, we present an $\{x,Q^2\}$ 
map of what is known and what is to come.

\section{PRECISION OF PDFS AND GLOBAL ANALYSES}

The extraction of PDFs from measurements is a complex process, involving
information from different experiments and a range of phenomenological and 
theoretical input.

   \subsection{Experimental systematic errors}

Since the extraction of the PDFs usually requires using data from
different experiments, and since the most precise data are usually 
limited by systematic, rather than statistical, errors, it is 
important that the systematic errors are taken properly into account.
In particular, it is necessary to understand the correlations of different
systematic errors on the measurements within and across experiments.  Several
groups have begun to make this information available in electronic and tabular
form.  Contributions to these proceedings by Tim Bolton (NuTeV) and 
Allen Caldwell (ZEUS) give the details.

   \subsection{$\{x,Q^2\}$ Kinematic Map for PDFs}
\def\figdis{
\begin{figure}[t]
\begin{center}
 \leavevmode
  \epsfxsize=3in  
  \epsfbox{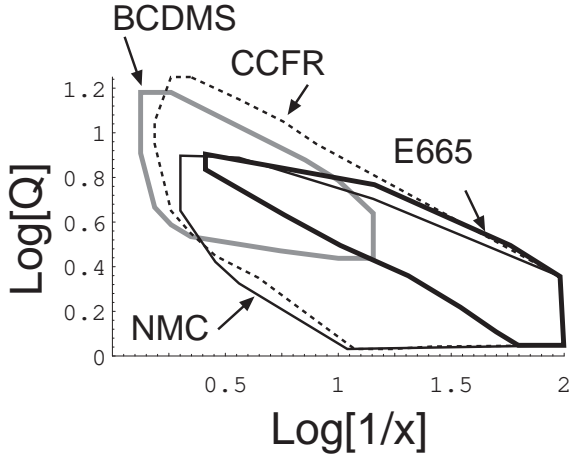}
 \end{center}
      \caption{Fixed target DIS data.  Note the full $\{x,Q^2\}$ region
               is clipped by the plot.
 }
   \label{fig:dis}
\end{figure}
}
\def\figdph{
\begin{figure}[t]
\begin{center}
 \leavevmode
  \epsfxsize=3in  
  \epsfbox{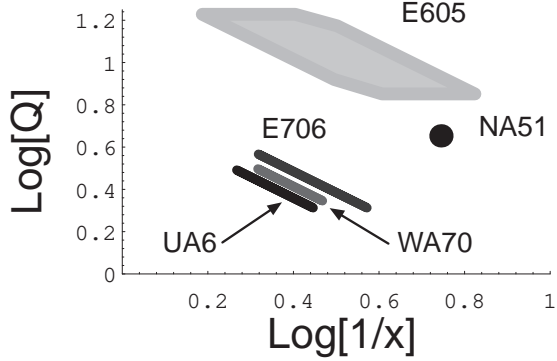}
 \end{center}
      \caption{Drell-Yan (E605), Direct Photon (E706, WA70, UA6),
and DY asymmetry (NA51) data.
 }
   \label{fig:dph}
\end{figure}
}
\def\fighera{
\begin{figure}[t]
\begin{center}
 \leavevmode
  \epsfxsize=3in  
  \epsfbox{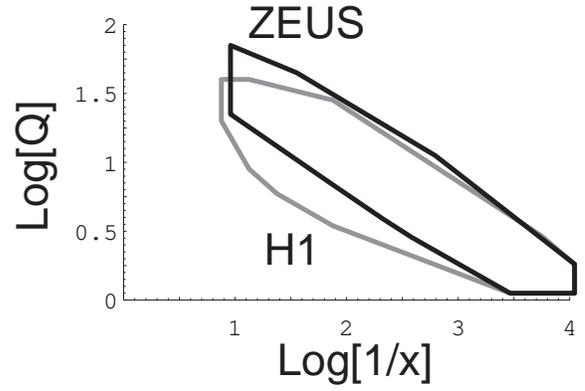}
 \end{center}
      \caption{$ep$ collider data.  Note the full $\{x,Q^2\}$ region
               is clipped by the plot.
 }
   \label{fig:hera}
\end{figure}
}
\def\figtev{
\begin{figure}[t]
\begin{center}
 \leavevmode
  \epsfxsize=3in  
  \epsfbox{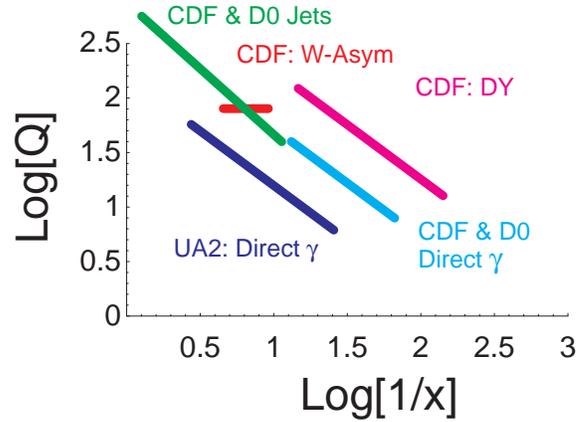}
 \end{center}
      \caption{Hadron-hadron collider data.
 }
   \label{fig:tev}
\end{figure}
}
\newcommand{\figGluAB}
{
\begin{figure}[hbt]
\epsfxsize=\hsize
\centerline{\epsfbox{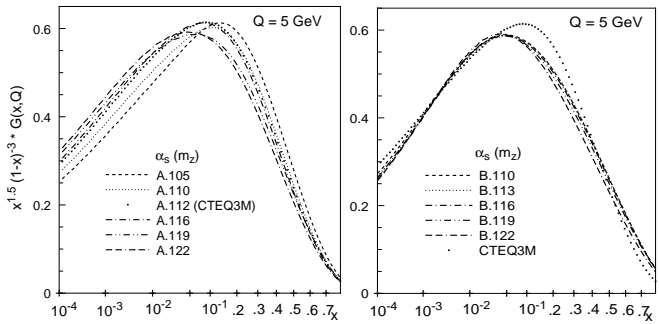}}
  \caption{Comparison of gluons obtained with pre-1995 DIS
data (A-series) with those using current DIS data (B-series). 
(Cf., Ref.~\protect\cite{cteq4}.)
 }
  \label{figGluAB}
\end{figure}
}

\newcommand{\figGluCdA}
{
\begin{figure}[hbt]
\epsfxsize=\hsize
\centerline{\epsfbox{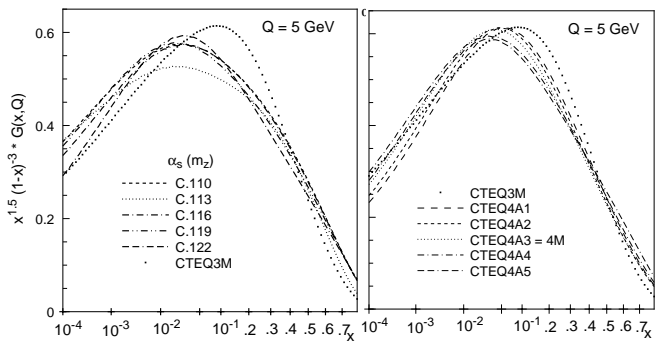}}
  \caption{Comparison of gluons obtained without jet
data (C-series) with those obtained with jet data, D-series (CTEQ4Ax).
(Cf., Ref.~\protect\cite{cteq4}.)
}
  \label{figGluC4A}
\end{figure}
}

 \figdis
 \figdph
 \fighera
 \figtev

 \figGluAB 
 \figGluCdA

Global QCD analysis of lepton-hadron and hadron-hadron  processes has made
steady progress in testing the consistency of perturbative QCD (pQCD)
within many different sets of data, and in yielding increasingly detailed
information on the universal parton 
 distributions.\footnote{PDF sets are available via WWW on the CTEQ page at 
http://www.phys.psu.edu/$\sim$cteq/ and 
on the The Durham/RAL HEP Database at 
http://durpdg.dur.ac.uk/HEPDATA/HEPDATA.html.
}
 We present a detailed compilation of the kinematic ranges covered by
selected experiments from all high energy processes
relevant for the determination of the universal parton
distributions.  This allows an overall view of the overlaps and the
gaps in the systematic determination of parton distributions;  hence,
this compilation
provides a useful guide to the planning of future experiments and to
the design of strategies for global analyses.

These analyses incorporate diverse data sets including
  fixed-target  deeply-inelastic scattering (DIS) data\cite{one,DisExp} of 
  BCDMS, CCFR, NMC, E665; 
  collider DIS data  of H1, ZEUS; 
  lepton pair production (DY) data of E605, CDF; 
  direct photon data of E706, WA70, UA6, CDF; 
  DY asymmetry data of NA51; 
  W-lepton asymmetry data of CDF; and
  hadronic inclusive jet data\cite{JetExp} of CDF and D0. 
 The total number of data points from these experiments is $\sim 1300$,
and these cover  a wide  region in the kinematic $\{x,Q^2\}$ space
anchored by HERA data at small $x$  and Tevatron jet data at high $Q$.

We now present the various experimental processes.
Note that while this is a comprehensive selection of
experiments, it is by no means exhaustive; we have attempted
to display those data which are characteristic for the
structure function determination.
In some cases, we have taken
the liberty to interpret the data so as to facilitate comparison among
the  diverse processes we 
 consider.\footnote{In particular, 
since we have taken the data points from the
global fitting files, there is 
 a cut on the minimum value of $Q \sim 2 GeV$ to avoid the non-perturbative region.}
 Also note that we have not attempted to deal with the different precision
of different measurements, or to separately consider the quark and gluon
determination; the reader should keep these points in mind when comparing
the figures.


 The quark distributions inside the nucleon have been quite well
determined from precise DIS and other processes, cf., Fig.~\ref{fig:dis}:
\begin{equation}
\mu,\nu   + N  \rightarrow \mu,\nu  + X 
\ .
\end{equation}
  Improved DIS data in the small-$x$ region is available from HERA, and
this is of sufficient precision to  be sensitive to the indirect influence
of gluons via high order processes.

The Drell-Yan process is related by crossing to DIS.
In lowest order QCD it is described by quark-antiquark annihilation:
\begin{equation}
q + \bar{q} \rightarrow \gamma^* \rightarrow \ell^+  + \ell^-  
\ .
\end{equation}
The kinematic coverage is shown in  Fig.~\ref{fig:dph} and  Fig.~\ref{fig:tev}.

Recent emphasis has focused on the more elusive gluon distribution, 
$G(x,Q)$, which  is  strongly coupled to the
measurement of $\alpha_s$.
 Direct photon production, 
 \begin{equation}
 g + q \rightarrow \gamma + q
 \qquad , \qquad
 q + \bar q \rightarrow \gamma + g
 \quad ,
 \end{equation}
in particular from the high statistics fixed target
experiments,  has long been regarded as the most useful source of
information on
$G(x,Q)$, cf., Fig.~\ref{fig:dph}. 
 However, there are a number of large theoretical uncertainties (e.g.,
significant scale dependence, and $k_T$ broadening of initial state
partons due to gluon radiation)\cite{CtqDph,TungJet} that need to be brought
under control before direct photon data can place a tight constraint on the
gluon distribution.

For example, the $k_T$ broadening due to soft gluon radiation is
essentially a higher twist effect (but with a large coefficient), and
should affect all hard scattering cross sections. The magnitude of the
correction to the cross section should be on the order of 
$n(n+1) \langle k_T\rangle^2/(4
p_T^2)$, where $\langle k_T \rangle$ is the average $k_T$ in the hard
scatter, and $n$ is the exponent of the differential cross section with
respect to $p_T:$ ($d\sigma/dp_T \propto 1/p_T^n$). For the Tevatron
collider regime, the effect should fall off as $\sim 1/p_T^2$, as is
observed for example in direct photon production in CDF. For $p_T > 50\, 
GeV$, the effect is negligible. For fixed target experiments, the
effective value of $n$ is large and changes rapidly with $p_T$ (due to the
rapidly falling parton distributions). The soft gluon radiation tends to
make the cross section steeper at low $p_T$ and at high $p_T$, and to cause an
overall normalization shift of a factor of 2.\cite{CtqDph}
        There are several approximate methods to predict the effects of
soft gluon radiation, as for example in gaussian $k_T$ smearing, or the
incorporation of parton showers into a NLO Monte Carlo.  Further
understanding may await the development of a more formal treatment of the
effect. Several theoretical ideas are under development.

Inclusive jet production in hadron-hadron collisions,
\begin{equation}
\{g g,\  q \bar{q}\} \rightarrow \{g g,\  q \bar{q}\}
\quad , \quad
g+\{ q,\   \bar{q}\} \rightarrow g+\{ q,\   \bar{q}\}
\ ,
\end{equation}
 is very sensitive to
$\alpha_s$ and $G(x,Q)$, (Fig.~\ref{fig:tev}).  NLO  inclusive jet cross
sections yield  relatively small $\mu$ scale dependence  for  moderate to
large
$E_t$ values.\cite{JetTh}
 High precision data on single jet production is now available over a wide
range of energies, $15\, GeV <E_t<450\, GeV$.\cite{JetExp}
 For $E_T > 50 GeV$, both the theoretical and experimental systematic
errors are felt to be under control.
 Thus, it is natural to incorporate inclusive jet data in a global QCD
analysis.

 
In reviewing the figures  we see the large kinematic range which is
explored by these processes.
 It is a useful exercise to overlay the  curves according to the separate
determination of the valence-quarks, light-sea-quarks, heavy-quarks, and
gluons. Although there is no room here for such a presentation, we leave
this as an exercise to the interested reader.
 Obviously, when comparing such a wide range of processes, one must  keep
in mind considerations beyond just the kinematic ranges. For example,  the
DIS and Drell-Yan processes are useful in determining the quark
distributions, whereas the direct photon and photoproduction experiments
yield information about the gluon distributions--though {\it not} with
comparable accuracy; the determination of the  gluon  distribution is
subject to many more theoretical and experimental uncertainties.

Likewise, the systematics for hadron-hadron and lepton-hadron
processes are quite different.
 Specifically,  while the hadron-hadron colliders can in principle
determine parton distributions out to large $Q^2$, extractions of PDFs
from these data are only beginning.
 DIS experiments  probe small $x$ (HERA)
and high $x$ (NuTeV), and low-mass 
Drell-Yan collider measurements  yield complementary results at higher
$Q^2$.  This combination of experiments  improves the reliability of the PDFs,
allows for cross checks among the different experiments, and yields precise
tests of the QCD evolution of the parton distributions.

\subsubsection{Progress of PDFs}

As new global PDF fits are being updated and improved, it can be difficult to quantify
our progress as to how precisely we are measuring the hadronic structure. 
To illustrate this progress we consider sets of global PDF fits using
various subsets of the complete data 
 set.\footnote{For the details of how these fits were performed,
see the original paper, Ref.~\cite{TungJet}.}

 First, we compare the A- and B-series of fits shown in Fig.~\ref{figGluAB}. 
The A-series shows a selection of gluon PDFs extracted from pre-1995 DIS
data using various values of $\alpha_s(M_Z^2)$ as indicated in the figure. 
 The B-series shows the same selection, but including the recent DIS data. 
By comparing the A- and B-series of
fits, we found that recent DIS data
\cite{DisExp} of NMC, E665, H1 and ZEUS considerably narrow
down the allowed range of the parton distributions. 

 Next, we compare the B- and C-series of fits shown in Fig.~\ref{figGluAB} and
Fig.~\ref{figGluC4A}. 
These fits were performed with the same data set, but the C-series fit used
a more generalized parametrization with additional degrees of freedom. 
By contrasting the B- and C-series we see that we must be careful to ensure
that our parameterization of the initial PDFs at $Q_0$ is not restricting the 
extracted distributions. 

Finally,  we compare the C- and D-series (CTEQ4Ax) of fits shown in Fig.~\ref{figGluC4A}. 
For the D-series fits, the Tevatron jet data was used, whereas this was
excluded from the C-series fits. 
 The jet data has a significant effect in more
fully constraining $G(x,Q)$  as  compared to the C-series.
 The quality of the final D-series fits (CTEQ4Ax) is indicative of  
the progress that has been made in this latest round
of global analysis.

   \subsection{High $E_t$ Jets and Parton Distributions} 
\newcommand{\figJetFit}
{
\begin{figure}[htbp]
\epsfxsize=\hsize
\centerline{\epsfbox{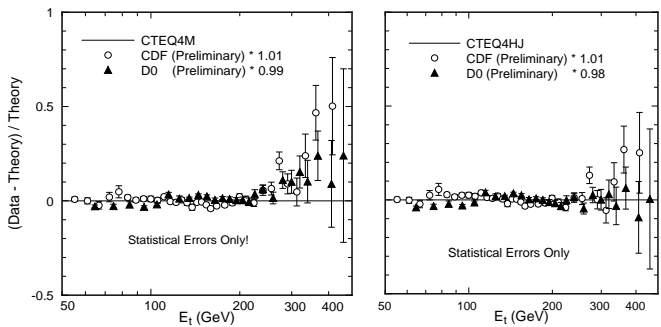}}
  \caption{CDF and D0 data compared to NLO QCD using a) CTEQ4M and b) CTEQ4HJ.
 {\it Cf.}, Refs.~\protect\cite{TungJet,cdfIa,cdfIb,d0Iab}.
}
  \label{figJetFit}
\end{figure}
}
\newcommand{\figHiEtGluon}
{
\begin{figure}[htbp]
\epsfxsize=3.5in
\centerline{\epsfbox{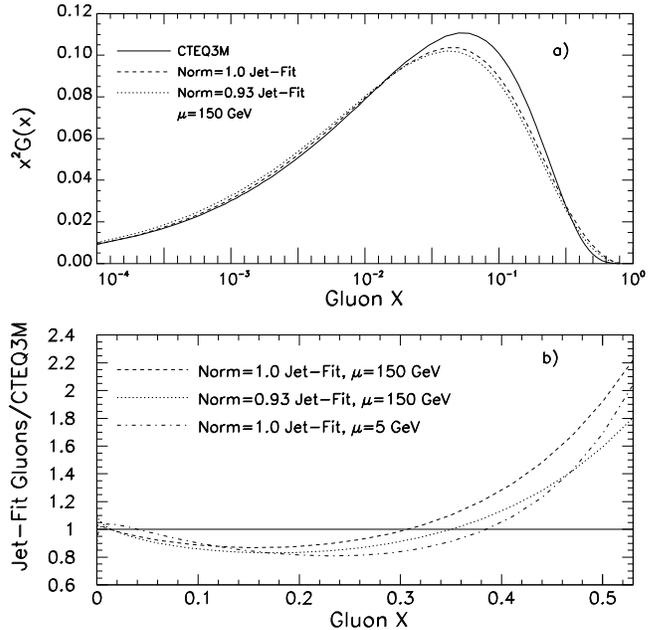}}
	\caption{(a) The CTEQ4HJ gluon distributions are compared to that of
CTEQ3M: (b) the ratio of the CTEQ4HJ gluons to 
 CTEQ3M.  {\it Cf.}, Ref.~\protect\cite{TungJet}. 
}
	\label{fig:higl150}
\end{figure}
}
\newcommand{\tblChiSq}
{
\begin{table}[htbp]
\begin{center}

\caption{Total $\chi^2$ ($\chi^2/point$) values and their distribution among
the DIS and DY experiments for CTEQ4M and CTEQ4HJ.
{\it Cf.}, Ref.~\protect\cite{cteq4}. 
}

\vskip 10pt

\label{tblChiSq}
\begin{tabular}{|c|c||c|c|c|}
\hline
 Experiment     & \#pts &  CTEQ4M     &  CTEQ4HJ     \\
\hline\hline
DIS-Fixed Target   &   817  &  855.2(1.05) &  884.3(1.08) \\ \hline
DIS-HERA   &   351  &  362.3(1.03) &  352.9(1.01) \\ \hline
DY rel.    &   129  &  102.6(0.80) &  105.5(0.82) \\ \hline
\hline
 Total     &  1297  &   1320      &   1343        \\ \hline
\end{tabular}
\end{center}
\end{table}
}

 \figJetFit 
 \figHiEtGluon  
 \tblChiSq

High statistics inclusive jet production measurements at the Tevatron
have received much attention recently because the high jet $E_t$
cross-section\cite{cdfIa,cdfIb,d0Iab} is larger than expected from NLO
QCD calculations.\cite{JetTh} A comparison of the inclusive jet data
of CDF and D0 and results is given in Fig.~\ref{figJetFit}a.  We see
that there is a discernible rise of the data above the fit curve
(horizontal axis) in the high $E_t$ region.  The essential question is
whether the high $E_t$ jet data can be explained in the conventional
theoretical framework, or require the presence of ``new
physics''.\cite{mrsd0,GMRS,TungJet}

Although inclusive jet data was included in the global fit of the PDF,
it is understandable why the new parton distributions (e.g.,
CTEQ4M)	still underestimate the experimental cross-section: these data
points have large errors, so they do not carry much statistical weight
in the fitting process, and the simple (unsigned) total $\chi^2$ is
not sensitive to the pattern that the points are uniformly higher in
the large $E_t$ region.  A recent study investigated the feasibility
of accommodating these data in the conventional QCD framework by
exploiting the flexibility of $G(x,Q)$ at higher values of $x$ (where
there are few independent constraints), while maintaining good
agreement with other data sets in the global analysis.\cite{TungJet}

A result of this study is the CTEQ4HJ parton sets which are tailored
to accommodate the high $E_t$ ($>200$ GeV) jets,\footnote{This set is
tailored to accommodate the high $E_t$ jets by artificially decreasing
the errors in the fit.  See Ref.\cite{TungJet} for details. The
$\chi^2$ of Table~\ref{tblChiSq} is computed using the true errors.}
as well as the other data of the global fit.\cite{TungJet}
Fig.~\ref{figJetFit}b compares predictions of CTEQ4HJ with the results
of both CDF and D0.\footnote{For this comparison, an overall
normalization factor of 1.01(0.98) for the CDF(D0) data set is found
to be optimal in bringing agreement between theory and experiment.}
Results shown in Fig.~\ref{fig:higl150} and Table~\ref{tblChiSq}
quantifies the changes in $\chi^2$ values due to the requirement of
fitting the high $E_t$ jets.  Compared to the best fit CTEQ4M, the
overall $\chi ^2$ for CTEQ4HJ is indeed only slightly
higher.\cite{cteq3,cteq4} Thus the price for accommodating the high
$E_t$ jets is negligible.

The much discussed high $E_t$ inclusive jet cross-section has been
shown to be compatible with all existing data within the framework of
conventional pQCD {\it provided} flexibility is given to the
non-perturbative gluon distribution shape in the large-$x$ region.

Presently, we note that the direct photon data from the Fermilab
experiment E706 are sensitive to the same $x$ range that affect the
Tevatron high $E_t$ jet data. A more quantitative theoretical
treatment of soft gluons may allow the direct photon data to probe
this question more precisely.  We will need such accurate, independent
measurements of the large-$x$ gluons to verify if the high-$E_t$ jet
puzzle is resolved, or whether we have only absorbed the ``new
physics" into the PDFs.

Nevertheless, this episode provides an instructive lesson: the
precision with which we know the PDFs is not indicated from a simple
comparison of different global fit sets.  These fits proceed from
similar assumptions and procedures, so their relative agreement should
not be taken as assurance of our knowledge of the PDFs.  In the
present case, the gluon density was naively estimated to be less than
10-20\% (in the $x$ kinematic range relevant for high $E_t$ jet
production) from a simple comparison of different PDF sets.
Surprisingly, a large change was eventually required (and accommodated)
by the data (assuming the Tevatron result is not an indication of some
new physics).
\subsection{Challenges for Global Fitting}

Global fitting of PDFs is a highly complex procedure which 
is both an art and a science. This requires fitting a large number
of data points from diverse experiments with differing systematics. 
Furthermore, the data are compared with NLO theory which introduces 
additional complications on the theoretical side.  

There was extensive discussion as to how
to determine the uncertainty of the PDFs.  We note that one of the
most important uncertainties for the PDFs is the choice of 
$\alpha_S$, since this affects the gluon distribution directly
as well as the singlet quarks.  Both MRS and CTEQ now provide
different PDFs with different choices of $\alpha_S$,  this is a big
improvement in determining the uncertainty of PDFs.  But this
group did not succeed in answering all the questions related to the
goal of a true one-standard-deviation covariance matrix of 
PDF uncertainties, although we did focus on some points that 
deserve further study.   We list some of these below.

\begin{enumerate}

\item
 A reminder:  when adding two experiments, you simply add their
$\chi^2$'s, and $\Delta\chi^2=1$ of the total $\chi^2$ is one
$\sigma$.

   \begin{enumerate}
   \item 
   Due to direct photon theory $\mu$-scale and $k_T$ 
   uncertainties, there is no way to define one standard deviation
   for these data.  The handling of the $\mu$-scale is done differently
   in different groups and can lead to somewhat different gluon distributions.

   \item 
   Other ``choices'' can lead to significant 
   differences in $\chi^2$ ($\Delta \chi^2 \approx 50-100$ units is typical). 
   These choices include which data sets to use, the starting $Q_0$ value, 
   etc.  One example is the small-x CCFR neutrino data which disagrees
   with the electron/muon DIS data.  This difference is unlikely 
   to be caused entirely by parton distributions,  and how this
   is handled in the global fits can cause significant changes
   in the global $\chi^2$.
   \end{enumerate}

\item
 Many experiments do not provide correlation matrices, and we've
never seen a correlation matrix for a theoretical uncertainty.
Without both of these for every experiment, one cannot expect
$\Delta\chi^2=1$ to work.

\item
 In principle we should add in LEP/tau/lattice
constraints on $\alpha_{S}$.   But if they are treated as only a
single data point, they will be swamped by the other 1200 points.
(This would not be true if $\Delta\chi^2=1$ were valid.)

\item
 What to do about the charm mass in DIS?  It will change $F_2$
predictions, but  the resulting parton distributions then
have a different definition of ``heavy quark in the proton'' and
this must be accounted for in the theoretical calculations.

\item
 When the CDF W asymmetry and NA51 data were added (the change
between CTEQ2 and CTEQ3), they gave a consistent picture
of $\bar{u}$ and $\bar{d}$.  But the $\chi^2$ went up for the rest of the
experiments by 30!  Once again $\Delta\chi^2=1$ is invalid.
The choice was to accept the larger $\chi^2$'s to incorporate the
new data presented.

\item
 What about higher twists?  Should a higher twist theoretical
uncertainty be added to DIS data?

\end{enumerate}

   \subsection{Choice of Parametrization} \label{subsec:parm}
\hyphenation{ha-dro-pro-duc-tion re-nor-mal-i-za-tion ex-ci-ta-tion }
\def\gsim{\, \lower0.5ex\hbox{$\stackrel{>}{\sim}$}\, }
\def\lsim{\, \lower0.5ex\hbox{$\stackrel{<}{\sim}$}\, }
\def\alphas{\alpha_S}

\def\figMRSg{
\begin{figure}[htbp]
\begin{center}
 \leavevmode
  \epsfxsize=3in
  \epsfbox{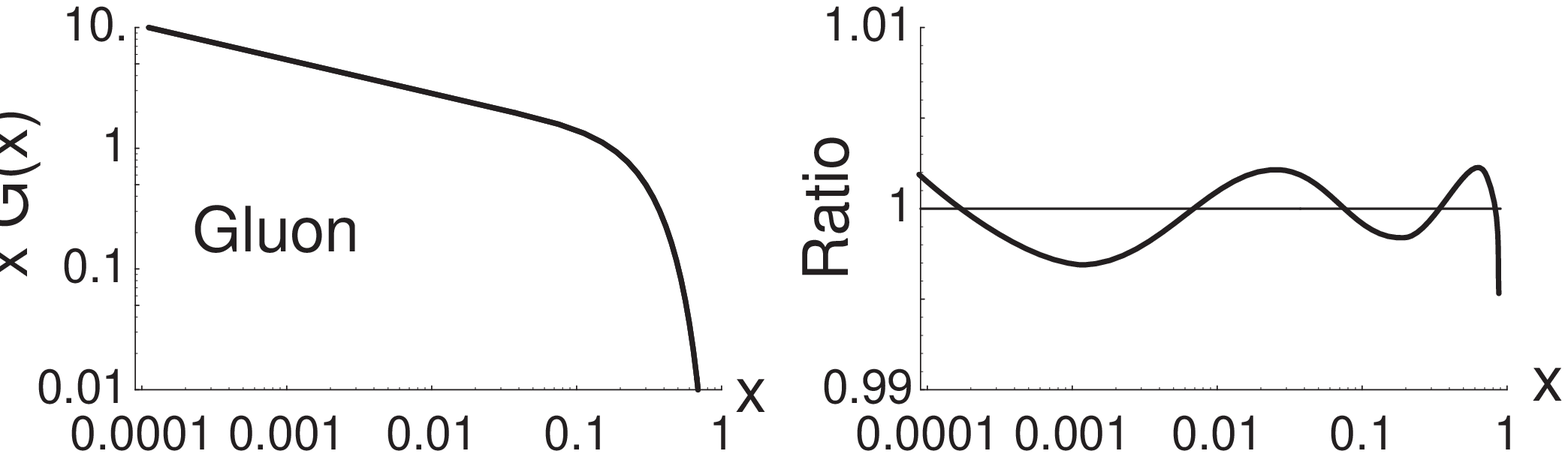}
 \end{center}
      \caption{a) The  gluon distribution $x G(x)$ at 
$Q_0 =1.6 GeV$ using the MRS and CTEQ parametrizations. 
The two curves are indistinguishable in this plot. 
b)  Fractional deviation for gluon of the
 CTEQ and MRS parametrizations. 
Note the full range of the y-axis is $\pm 1\%$. 
 }
   \label{fig:MRSg}
\end{figure}
}
\def\figMRSuv{
\begin{figure}[htbp]
\begin{center}
 \leavevmode
  \epsfxsize=3in  
  \epsfbox{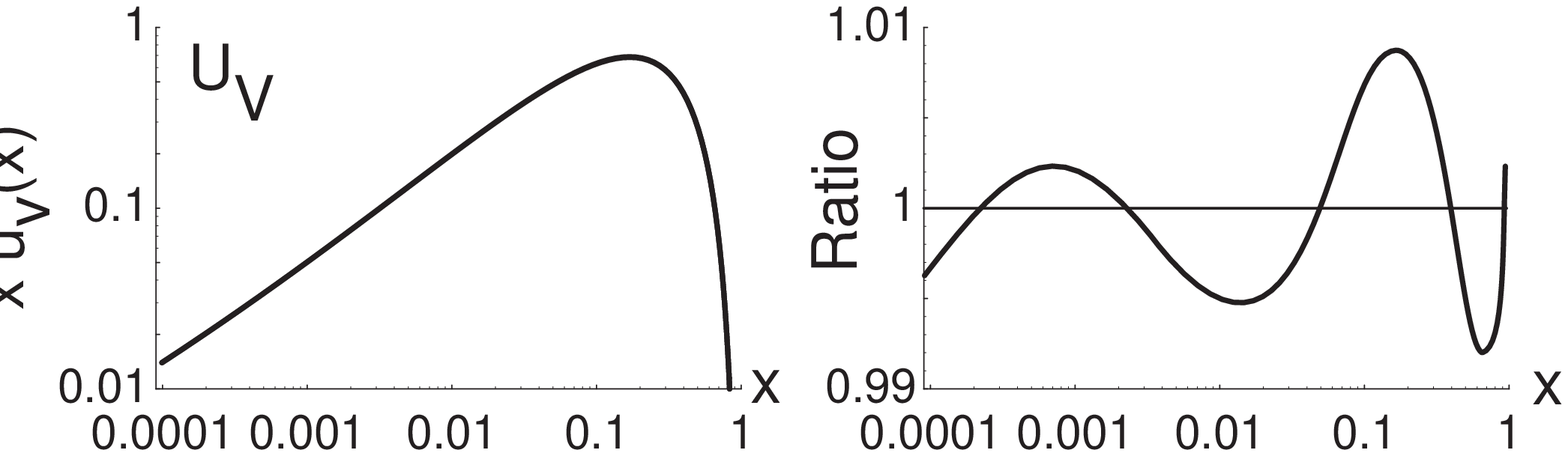}
 \end{center}
      \caption{a) The  u-valence distribution $x u_v(x)$ at 
$Q_0 =1.6 GeV$ using the MRS and CTEQ parametrizations. 
The two curves are indistinguishable in this plot. 
b)  Fractional deviation for u-valence of the
 CTEQ and MRS parametrizations. 
Note the full range of the y-axis is $\pm 1\%$. 
 }
   \label{fig:MRSuv}
\end{figure}
}
\def\figMRSdv{
\begin{figure}[htbp]
\begin{center}
 \leavevmode
  \epsfxsize=3in  
  \epsfbox{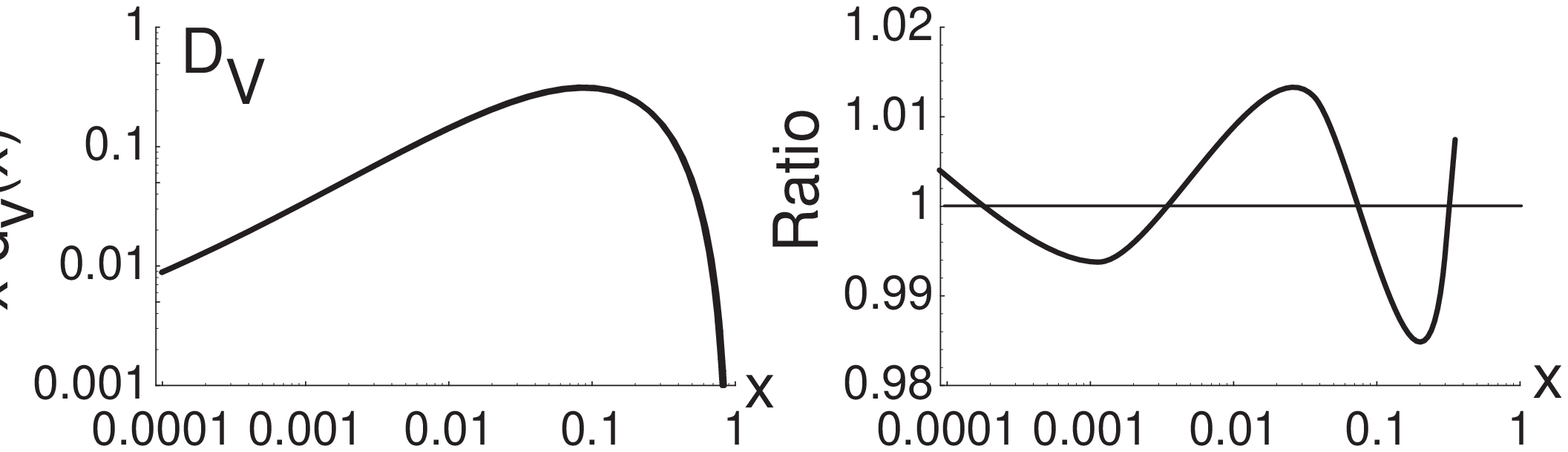}
 \end{center}
      \caption{a) The  d-valence distribution $x d_v(x)$ at 
$Q_0 =1.6 GeV$ using the MRS and CTEQ parametrizations. 
The two curves are indistinguishable in this plot. 
b)  Fractional deviation for d-valence of the
 CTEQ and MRS parametrizations. 
Note the full range of the y-axis is $\pm 2\%$. 
 }
   \label{fig:MRSdv}
\end{figure}
}
\def\figMRSev{
\begin{figure}[htbp]
\begin{center}
 \leavevmode
  \epsfxsize=3in  
  \epsfbox{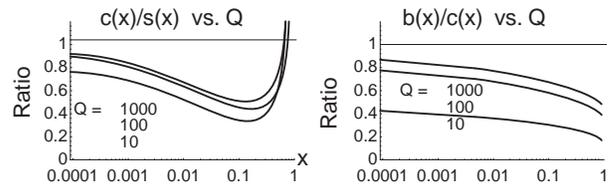}
 \end{center}
      \caption{The ratio of a) $c(x)/s(x)$ and b)  $b(x)/c(x)$ 
 for a range of $Q$. For increasing $Q$, the evolution reduces any
difference between the distributions. 
 }
   \label{fig:MRSev}
\end{figure}
}
  \figMRSg
  \figMRSuv
  \figMRSdv
  \figMRSev

The choice of boundary conditions for the PDFs at the initial $Q_0$
has received increasing attention as the accuracy of data
improves.\cite{cteq4,GMRS,cteq3,Yndurain} Although the DGLAP evolution
equation clearly tells us how to relate PDFs at differing scales, the
form of the distribution at $Q_0$ cannot yet be derived from first
principles, and must be extracted from data.  For this purpose, it is
practical to choose a parametrization for the PDFs at the initial
$Q_0$ with a small number of free parameters that can be fit to the data.

A question that was repeatedly raised in the workshop is the extent to
which the choice of parametrization limits the extracted PDFs of the
global fit.  It is important to note that the evolution equation for
the global fits is solved numerically on an $\{x,Q^2\}$ grid.
Therefore the issue of the parametrization is only relevant at $Q_0$.
For $Q>Q_0$, the parametrized form is replaced by a discrete
$\{x,Q^2\}$ grid.

To approach these questions in a quantitative manner, we performed a
simple exercise to examine the potential difference of PDFs that can
be attributed to different choices of parametrizations.
Specifically, we investigated the difference between the
MRS\cite{GMRS} and CTEQ\cite{cteq3} parametrizations, which take the
general form:

\noindent
MRS:
 \begin{equation}
 f(x,Q) = a_0 x^{a_1} (1-x)^{a_2} (1+a_3 \sqrt{x} + a_4 x)
 \label{eq:parm}
 \end{equation}
CTEQ:
 \begin{equation}
 f(x,Q) = b_0 x^{b_1} (1-x)^{b_2} (1+b_3  x^{b_4}) 
 \label{eq:parm2}
   \end{equation}
We used the CTEQ3M PDF set at $Q_0 = 1.6\, GeV$ (which is naturally
described by the CTEQ parametrization shown above), and performed a
fit to describe the same PDF set using the MRS parametrization.  Note
that this is an academic exercise that does not fit data, but rather
explores the flexibility of the parametrizations.

If we can accurately describe the CTEQ3 PDFs with the MRS form, then
it is plausible that the particular parametrization choice for the
PDFs has little consequence. However, if we cannot accurately
describe the CTEQ3 PDFs with the MRS form, we will need to
investigate more thoroughly whether the parametrization introduces a
strong bias as to the possible PDFs which will come from the global
fitting procedure.  

In Figs.~\ref{fig:MRSg}, \ref{fig:MRSuv}, and \ref{fig:MRSdv}, we plot
both the CTEQ3 PDFs and the fit to the CTEQ3 PDFs using the MRS
parametrized form, Eq.~\ref{eq:parm}.  We only show the gluon,
u-valence, and d-valence; the results for the sea-quarks will be
similar to the gluon.  First we plot $x\, f(x)$ at $Q_0=1.6\, GeV$ on
a Log-Log scale.  The two separate curves are indistinguishable in
this plot.  To better illustrate the differences, we plot the
fractional difference between the two PDFs.  Observing that the scale
on this plot is $\leq 2\%$, we see that the variation over the range
$x=[10^{-4},1]$ is relatively small.  We find larger deviations in the
high $x$ region, but the significance of this is diminished by the
fact that the PDFs are small in this region.

Although we do not claim that this is an exhaustive investigation,
this simple exercise appears to indicate that the PDFs extracted from
a global fit should be insensitive to the choice of the above
parametrizations (Eq.~\ref{eq:parm}).  One might speculate that the
same conclusion would also hold for other parametrizations; however,
such an exercise has yet to be performed.

Furthermore, since the QCD evolution is stable as one evolves up to
higher values of $Q$, any small differences at $Q_0$ will decrease for
$Q>Q_0$. We can roughly see this effect by examing the ratios of
$c(x)/s(x)$ and $b(x)/c(x)$ as shown in Fig.~\ref{fig:MRSev}. For
example, at $Q= 10\, GeV$, the b-quark is less than half of the
c-quark distribution; at $Q= 100\, GeV$ the b-quark distribution is
significantly closer to the c-quark.\cite{collinstung86} This
observation suggests that the small differences we observed at $Q_0 =
1.6\, GeV$ will quickly wash out as we evolve upwards.

The above observations, however, only apply in regions of $x$ where PDFs  
are well-determined; and they cannot be taken literally without qualification.
An important example which illustrates the importance of exercising caution  
is the behavior of the gluon distribution at large $x$ brought to focus by the
high $E_t$ jet data, as discussed in the last section. Whereas, using  
``conventional'' parametrization of the form Eq.(\ref{eq:parm}), 
GMRS \cite{GMRS} found it  
impossible to fit the jet results along with the rest of the global data, CTEQ
showed that allowing for a more flexible parametrization of the gluon  
distribution at large $x$ can accommodate both. To accomplish this, one will  
need a functional form such as Eq.(\ref{eq:parm2}), with $b_4$ substantially 
bigger than 1,
or equivalently, $e^{b_4x}$ in place of $x^{b_4}$. (Since $0<x<1$, and the  
whole expression is multiplied by $(1-x)^{b_2}$ which is steeply falling,  
$G(x,Q)$ is still well-behaved.) The difference in the size of $G(x,Q)$  
resulting from these parametrizations can be as much as 100\% at $x=0.5$, as  
shown in Fig.~\ref{fig:higl150}.

\section{STRUCTURE FUNCTIONS AND $\alpha_{S}$} \label{subsec:alphas}
     Structure functions are important in that they give us
information on the value of $\alpha_s$, and also in that they are
often inputs to many different measurements, some of which themselves
are used to determine $\alpha_s$.  In this summary of the work of the
joint $\alpha_s$-structure function groups we investigate
how structure functions themselves give us direct information on
$\alpha_s$, and the expected uncertainties of possible new measurements of
structure functions at future proposed machines.  There are two
categories of structure function analyses which result in an
$\alpha_s$ measurement: $Q^2$ evolution of structure functions, and
measurements of sum rules, which pertain to the integrals of specific
structure functions over $x$.  Since the theoretical and experimental
errors are comparable for some of these analyses, this report  
examines how improvements might be made in both areas.

     On the experimental side of the study, we consider a
$\mu p$ collider or an $e p$ collider, and also a neutrino scattering
experiment at a $\mu^+ \mu^-$
collider.  Given the current level of 
error in $\alpha_s$ measurements, we consider here only analyses which may
result in few per cent or less error on $\alpha_s(M_Z^2)$.  To address the
theoretical issues within the scope of this report we can at best point
out the largest problems and how they are currently being investigated, 
in hopes of inspiring theorists to devote more time to them.

\subsection{Evolution of Structure Functions}

     When looking at the $Q^2$ evolution of structure
functions, one can use the DGL-Altarelli-Parisi Equations to find
$\alpha_s$ \cite{dglap}.  
In the case of the non-singlet structure functions the evolution as a
function of $Q^2$ is simply related to $\alpha_s$ and the
non-singlet structure function itself.  In the case of the singlet
structure functions, the evolution is related to $\alpha_s$, the
structure function itself, AND the gluon distribution, which
complicates matters.  In either case care must be taken avoid large higher
twist effects, which are particularly important at low \qsq.  

Non-singlet structure functions can be measured in both
neutrino and charged lepton scattering experiments.  
One way is by taking the average of 
$xF_3^{\nu N}$ and $xF_3^{\overline \nu N}$, where $\nu N$ in the 
superscript indicates the presence of an isoscalar 
target.  Similarly, averaging $xW_3^{l^+d}$ and $xW_3^{l^-d}$ also results 
in a pure non-singlet structure function, where the lepton is either 
an electron or muon, and the scattering center is a deuteron.  Finally, 
one can use the structure function $F_2^{\nu N}$ or $F_2^{l^\pm N}$ at 
high x, since there are virtually no sea quarks at high $x$.  

Many high statistics determinations of $\alpha_s$ have to date been performed,
using a variety of techniques.  By fitting only $xF_3$ or $F_2$ at high 
$x$, one can do a pure non-singlet fit to the evolution, with no dependences
on the gluon distribution.  Given the wealth of precise data in charged lepton
scattering structure functions, however, there are also determinations of 
\alfs\ from fitting $F_2$ at all $x$, but including a contribution to the 
evolution from the gluons.  These two different kinds of determinations
do not show any systematic difference in the final result, as is shown 
in table \ref{tab:disresults}.  
\begin{table}[h]
\begin{center} 
\caption{A selection of $\alpha_s$ measurements from structure functions, and  
the total error on $\alpha_s(M_Z^2)$.} 
\label{tab:disresults}  
\begin{tabular}{lccl} 
Method & Experiment &\qsq\ & \alfs(\mztwo) \\ 
\hline\hline
$xF_3$ only & CCFR \cite{newccfr} & 25 & $.118\pm.007$ \\ 
$xF_3$ and $F_2$ & CCFR \cite{newccfr} & 25 & $.119\pm.0055$ \\ 
$F_2$ low $x$ & NMC \cite{nmc} &  7 & $.118\pm.015$ \\ 
$F_2$ high $x$ & SLAC$/$BCDMS \cite{virmil}& 50 & $.113\pm.005$ \\ 
$F_2$ low $x$   &  HERA \cite{hera}& 4-100 &  $.120\pm.010$ \\ 
\end{tabular} 
\end{center} 
\end{table} 
     The errors listed in \ref{tab:disresults} are deceiving, however,
because in fact they are all dominated by either experimental or theoretical
systematic errors.  In the remainder of this section we consider the 
largest two systematic errors, and how new machines (and new calculations) 
could hopefully reduce these errors.  

\subsubsection{Experimental Errors on \alfs\ and possible improvements} 
The dominant experimental systematic error in the measurements listed 
in the table come from energy uncertainties.  These
can come from spectrometer resolution, calibration uncertainty in the
detector, or overall detector energy scale.  The key to improving the
overall experimental error in these measurements is not higher
statistics or higher energies, but better calorimetry, and better
calibration techniques.  The challenge in determining the energy scale
in deep inelastic scattering experiments is in finding some ``standard
candle'' from which to calibrate.  For example, if there were some way
of measuring the known mass of some particle decaying in the system,
or if the initial beam energy was very well known because of
accelerator constraints, this could substantially improve the energy
scale determination over previous experiments.

     A number of machines have been proposed at this workshop in a 
variety of energies and initial particles.  While it is true that machines
(and experiments) are not proposed these days to do precise QCD 
measurements alone, there are some interesting possibilities that may arise 
from these machines.  

Because of other considerations (namely the rise of $F_2$ at low $x$)
a lepton/hadron collider is an attractive possibility.  Currently,
however, the HERA \alfs\ experimental error is dominated by
uncertainties in the $x$ distribution of the structure functions
measured, (particularly that of the gluon).  
In order to do a DGLAP-style evolution measurement in a
lepton hadron collider, one would need to have both $\ell^+p$ and
$\ell^-p$ collisions, measure the different cross sections, and
extract a non-singlet structure function.  The statistics needed for a
precise structure function measurement at the energies that have been
proposed would be well above current HERA expectations, and the higher
in \qsq\ these machines operate the lower the cross section, and the
smaller the effect one is trying to measure.

Another intriguing possibility would be a neutrino experiment at a
muon collider.  A 2TeV muon collider could (with considerable
engineering) make very high rate 800GeV neutrino and antineutrino
beams.  If one knew the muon beam energy very well (taking as an
example how well the LEP energy scale is now known after much work!)
then a neutrino beam coming from muon decays would be at a very
well-understood energy as well.  There would be negligible production
uncertainty from a neutrino beam coming from a muon beam, and the
rates for such a beam would be astronomical simply starting with the
current proposals for muon intensities in the accelerator.

\subsubsection{Theoretical Errors on \alfs\ and possible improvements} 

Currently the renormalization and factorization scale uncertainties
dominate the theoretical error on \alfs\ from structure function
evolution.  This is true for both singlet and non-singlet structure
functions evolution.  By assuming the factorization and
renormalization scales were $k_1\qsq$ and $k_2\qsq$ respectively, and
varying $k_1$ and $k_2$ between 0.10 and 4, Virchaux and Milsztajn
arrive at an error of $\delta(\alfs(\mztwo))= .004$.\cite{virmil}.
They claim that the overall $\chi^2$ of the fit did not increase
significantly when these variations were made.  Similar or larger QCD
scale errors apply to the other \alfs\ measurements listed in table
\ref{tab:disresults}.  Certainly the most straightforward (and perhaps
naive) way to reduce this error would be to calculate the next
higher-order term in the DGLAP equations.  

Still another method of reducing these errors is to actually fit for
$k_1$ and $k_2$, and see what the resulting error on these values are
within the fit.  By floating those constants, however, one is assuming
QCD works, and getting a good fit for one consistent value of
\lmsb\ in the experiment can no longer be claimed by itself as a test
of QCD.  If $k_1$ and $k_2$ are floated, one does not test QCD until
one compares one experiment's \alfs\ value with another experiment's
value.  Furthermore, if the fit prefers values of $k_1$ and $k_2$ far
away from one, one would also question the validity of the
measurement.

\subsection{Sum Rules are Better than Others }
     The two sum rules that have thus far been used to measure
$\alpha_s$ are the Gross Llewellyn Smith sum rule \cite{gls} and the
Bjorken sum rule \cite{bjsr}, which are related to $xF_3$, and the
polarized structure functions $g_n(x)$ and $g_p(x)$ respectively.
Since these methods of determining \alfs\ are far less mature than the
structure function analysis, the corresponding experimental errors on
\alfs\ are much larger. Since both Sum Rules come are fundamental
theoretical predictions, and the higher order corrections to the sum
rules are so straightforward to compute, the QCD scale error on 
these measurements is much smaller than those of the evolution
measurements.  Table \ref{tab:errglsbj} gives a list of systematic and
statistical errors for both sum rules.  

\begin{table}[h]
\caption{Table of errors on $\alpha_s(M_Z^2)$ from Sum Rules.
$^\dagger$ From E142 result, E154 claims should be higher} 
\label{tab:errglsbj} 
\begin{center} 
\begin{tabular}{|l|c|c|}
\hline 
   & \multicolumn{2}{|c|}{$\delta\alfs(\mztwo)$} \\ 
Error Source & GLS  & Bjorken \\ 
\hline 
Statistical & .004    & $<.001$         \\ 
\hline 
Low $x$ extrapolation  & .002  & .005$^\dagger$ \\ 
\hline 
Overall Normalization & .003 & .002 \\ 
\hline 
Experimental Systematics & .004   & .006  \\ 
\hline 
Higher Twist  &  .005   &  .003-.008 \\ 
\hline 
QCD Scale Dependence  &.001   & .002   \\ 
\hline   
\end{tabular} 
\end{center} 
\end{table} 

\subsubsection{Low $x$ Uncertainties} 
     The largest uncertainties in sum rule measurements come from the
fact that they involve integrals from $x=0$ to $x=1$.  Of course no
experiment can measure all the way down to $x=0$, and the closer to 
0 one can reach the smaller the error in extrapolating from the lowest
data point to zero will be.  What is usually done to extrapolate to
$x=0$ is a functional form is assumed, and the data are either fit to
that functional form and the resulting parameters are checked with a
theory, or if the data do not have enough statistical precision some
functional form is simply assumed.  While for the GLS sum rule the
data seem to agree with simple quark counting arguments for the form
of $xF_3$ at low $x$, the newest data from SLAC E154 (shown after
Snowmass'96 at ICHEP96) do not fit to a function whose integral
converges as $x$ goes to 0.  The collaboration does not yet
report a measurement of \alfs\ from their new data, saying that the low 
$x$ behavior of the integral is too uncertain; however,
this analysis is in progress.  For future improvement 
on the Bjorken Sum Rule one will need to go to lower $x$ than what has 
currently been reached ($x=.015$).  If the Bjorken integral is not finite
then much more is called into question than the validity of QCD!  

Another uncertainty associated with the low $x$ region is that one also
needs to go to low \qsq\ to measure low $x$.  At low \qsq\ higher twist 
uncertainties become important, and these higher twist contributions have 
never been measured for these sum rules.  The present state
of higher twist calculations for DIS sum rules is given in reference
\cite{bktwist}, which discusses results from many models of higher
twist calculations, including QCD Sum Rules, 
and a non-relativistic quark model.  Again, there is more trouble associated
with the Bjorken Sum Rule than the GLS sum rule, because the different 
models predict very different higher twist contributions to the former, 
while agreeing at the $50\%$ level for the latter.  So, whether one takes
as the higher twist error the spread of theoretical predictions or the 
error on one such prediction (shown in the table above) one can arrive 
at very different errors.  In either case that error is significant at 
the currently relevant \qsq\ region.   Unless a proven agreed-upon
method of higher twist calculations arises the best bet in the future
will be to simply fit the sum rule results for a higher
twist contribution and an $\alpha_s$ contribution. This will require
much higher statistical precision in the structure function
measurements themselves than what is currently available.  

\subsubsection{Normalization Uncertainties} 

\begin{figure}[b]
\leavevmode
\centerline{\epsfxsize=6cm \epsfbox{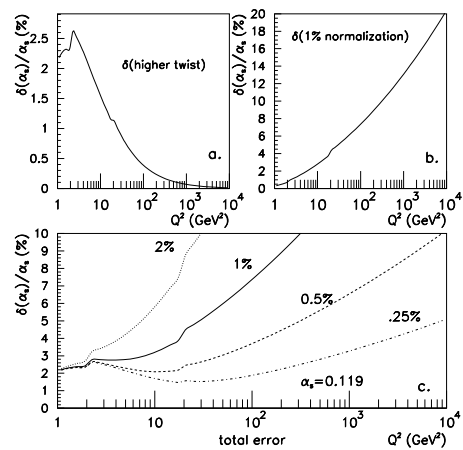}}
\caption{Variation of (a) higher twist, (b) $1\%$ normalization errors, 
and (c) the sum in quadrature of the two as a function of \qsq\ for the 
GLS sum rule.}
\label{fig:wheregls}
\end{figure}

Finally, if one proposes to improve these measurements by going to a 
higher \qsq\, the next most important error (assuming one has solved the 
problem of extrapolating to low $x$) will be the overall normalization 
error.  Since the effect one is measuring is
proportional to 1-\alfs and not \alfs, as \qsq\ gets larger and 
\alfs\ gets smaller then an
overall $1\%$ error on the normalization of the structure functions (and
hence the integral itself) turns into a larger fractional error on 
\alfs.  This is shown
quantitatively in figure \ref{fig:wheregls}, which shows the effects
of the higher twist error as a function of \qsq\ and a $1\%$
normalization error on the structure functions as a function of \qsq .
The sum of the two errors in quadrature show that measuring the sum
rules at a \qsq\ above 100\,GeV$^2$ will not reduce the overall error
for even an ambitious normalization error of $0.25\%$.  The current
normalization error on the overall $\nu$-nucleon cross section is
$1\%$, and the error on the ratio of $\nu$ and ${\overline \nu}$ cross
sections is another $1\%$, which translates into presently a total
$xF_3$ normalization of $1.4\%$.  There are currently no plans to
improve this measurement, one would need a tagged neutrino beam (which
might be possible with a muon beam at a muon collider) to do so.
 
\subsection{\alfs\ conclusions } 

     Structure functions and QCD provide us with the possibility 
of two complementary measurements of \alfs; the \qsq\
evolution and sum rules.  The current errors on \alfs\ from structure function 
evolution are in the $4-5\%$ range, and will be improved only with 
a reduction of the renormalization and factorization scale uncertainties.  
For this, next to next to leading order (NNLO) corrections to the 
DGLAP equations must be computed.  By far the most important experimental 
uncertainty in evolution measurements comes from how precisely experiments 
know their energy scale and resolution.  Sum Rules have very different 
outstanding issues, namely the low $x$ uncertainty and measurement, and 
also the higher twist terms.  The best way to eliminate higher twist 
uncertainties would be to simply measure their contributions in the 
lowest \qsq\, yet still have enough statistics at higher \qsq\ for a 
measurement of \alfs.  While the sum rule analyses would benefit from 
much higher statistics, in general, to arrive at new measurements of \alfs\ 
from structure functions we must do more than simply raise the energies of 
the experiments and run them longer!

\section{STRUCTURE FUNCTION INPUTS TO PRECISION ELECTROWEAK MEASUREMENTS}
Structure functions are inputs to many precision electroweak
measurements--a few examples are from $sin^2\theta_W$ measured in
$\nu$N scattering and global electroweak fits which include $\alpha_s$
from structure function data along with other fundamental parameters.
A measurement expected to have significant experimental improvement
in the future such that the structure function uncertainty becomes 
important relative to other uncertainties is the W
mass ($M_W$) measurement from on-shell production at collider experiments.  
Even at the current level of precision of this analysis there are 
outstanding questions about how that uncertainty is evaluated, and 
whether this could be improved, even before new experiments come around.  

At a hadron collider experiment, the W mass itself cannot be directly 
measured on an event by event basis, because the clean signatures 
of $W$ production contain a charged lepton and therefore also contain a
neutrino.  Furthermore, the initial center of mass 
energy of the partons which interact to give a $W$ is not known, so one 
cannot simply require the total momenta to balance to give the energy 
of the outgoing neutrino.  One can use the constraint that the total 
initial transverse momentum is zero, however.  The way the mass is then 
measured in an experiment is that the transverse mass is computed, 
$M_T=\sqrt{2p_t^\ell p_t^\nu (1-\cos \phi^{\ell \nu})}$, where 
$p_t^{\ell,\nu}$ are the transverse momenta of the charged lepton and 
neutrino, and $\phi^{\ell \nu}$ is the angle between the charged lepton
and neutrino in the transverse plane.  The shape of the $M_T$ distribution
is then extremely sensitive to $M_W$, but is also dependent on the 
parton distributions used in the Monte Carlo simulation, in particular, 
the transverse component of $u-d$.  

Table \ref{tab:werr} gives the uncertainty in $M_W$ from CDF and D0 from 
direct production, and measurement of the transverse mass \cite{youngkee}.  
\begin{table}[b] 
\caption{Table of uncertainties for both the CDF and D0 $W$ mass 
measurements, for different final states ($e\nu$ or $\mu\nu$) and 
different run periods (Ia,Ib).}  
\label{tab:werr}
\begin{tabular}{|l||cc||c|c||} \hline\hline 
Source & \multicolumn{2}{c}{CDF} & D0 & D0 \\ \hline 
  & \multicolumn{2}{c}{Ia} & Ia & Ib \\ 
  & $e$ & $\mu$ & $e$ & $e$ \\ \hline\hline 
Statistics & 145 & 205 & 140 & 70 \\ \hline 
Lepton Scale & 120 & 50 & 160 & 80 \\\hline 
Lepton Resolution & 80 & 60 & 85 & 50 \\ \hline
Lepton Efficiency & 25 & 10 & 30 & 20 \\ \hline
$P_T^W$, PDF & 65 & 65 & 65 & 65 \\ \hline
$P_T^{\rmt Recoil}$ Model & 60 & 60 & 100 & 55 \\ \hline
Underlying Event & & & & \\
in Lepton Towers & 10 & 5 & 35 & 30 \\ \hline
Background & 10 & 25 & 35 & 15 \\ \hline
Trigger Bias & 0 & 25 & - & - \\ \hline
QCD Higher Order Terms & 20 & 20 & - & - \\ \hline
QED Radiative Corrections & 20 & 20 & 20 & 20 \\ \hline
Luminosity Dependence & - & - & - & 70 \\ \hline\hline 
Total & \multicolumn{2}{c}{180} & 270 & 170 \\ \hline\hline 
\end{tabular} 
\end{table} 

\begin{figure}[tbph]
\vspace{6cm}
\includegraphics{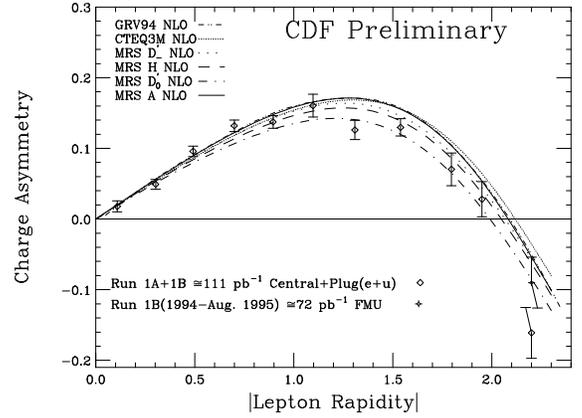}
\vspace{1cm}
\caption{The W asymmetry as measured in CDF and the prediction of various 
different parton distribution functions.}
\label{fig:wasym}
\end{figure}

Currently the structure function uncertainty is estimated by doing the
analysis with several different sets of parton distribution functions,
and comparing the results, using the W asymmetry measurement as
another constraint.  Figure \ref{fig:wasym} shows the measured W 
asymmetry from CDF and the predictions from various PDFs \cite{wasym}.  
Given that most of these PDFs come from the same input data (deep inelastic 
structure functions), the spread of the 
predictions represents an error in the technique of parametrizing the 
distribution which accounts for the W asymmetry, not the error on 
the distribution itself.  By requiring a PDF to reproduce the measured
W asymmetry, one is choosing a more appropriate parametrization, but one 
must go further to assign errors on that specific parametrization.  

\begin{figure}[th]
\vspace{6cm}
\includegraphics{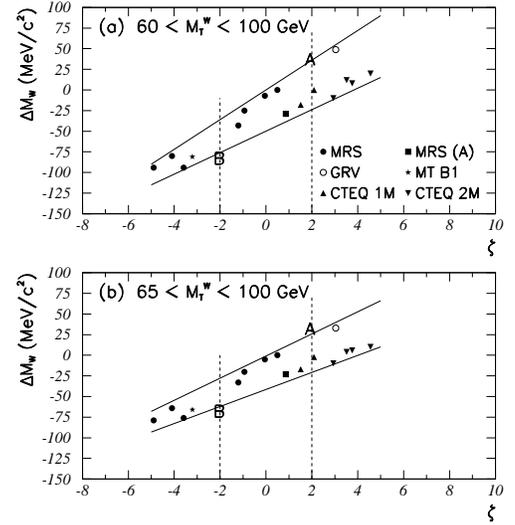}
\vspace{1cm}
\caption{The change in W mass versus the signed standard deviation 
of agreement with the measured W charge asymmetry for different PDFs.}
\label{fig:wmass_pdferr}
\end{figure}

Figure \ref{fig:wmass_pdferr} shows the resulting
change in $M_W$ for different PDFs, and how many standard deviations 
each PDF is from predicting the $W$ asymmetry \cite{wmass}.   

The problem with estimating this uncertainty by comparing different 
PDFs is the following:  if all of these PDFs are simply different 
parametrizations which come from the same sets of deep inelastic 
scattering data, then two different PDFs do not necessarily encompass
the uncertainty on whatever quark distributions are relevant.  There
must be errors on the PDFs in order for the correct error on the 
W mass uncertainty to be evaluated.  Of course, since at the present 
time there are no errors given with PDFs, this is not possible. 

There was much discussion at Snowmass about the difficulties associated
with assigning errors to PDFs, and one should refer to that section 
of this write-up, and a separate submission by Tim Bolton on this
topic.  Given that the job of assigning those errors is one that is 
far from completion, a temporary solution was suggested at this 
meeting.  Namely, a PDF-generator could produce a set of PDFs 
that span the range of the possible values of the distribution in 
question.  For example, for the jet $E_T$ analysis, a set of PDFs 
with different values of \alfs\ has been generated.  Similarly, 
a set of PDFs with the acceptable range of $u-d$, which is important
for the W mass measurement ( and also the W asymmetry measurement) could 
also be generated.  Then the $M_W$ analysis could simply compare the 
different PDFs in one set provided for an estimate on the $M_W$ error 
from uncertainty in the PDFs, and compare different parametrizations 
for the uncertainty on the parametrization.  

Furthermore, much care must taken when
using the measured $W$ asymmetry to constrain the $W$ mass error from
PDFs.  Since the $W$ mass is measured using a 
distribution dependent mostly on transverse quantities, 
and the asymmetries depend on longitudinal differences
between the $u$ and $d$ quark distributions, the correlations (and/or
lack of correlations) must be taken into account appropriately.

In order to use PDFs to their full potential, and also make precision
measurements at hadron collider experiments, collaboration between 
PDF-generators and experimenters is essential.  The $W$ mass illustrates
where this would be useful probably better than any other precision 
electroweak measurement.  Given that the future seems to be evolving towards
higher energy hadron colliders, the necessity of errors on parton distribution 
functions can only increase, as will the care required in using these 
functions correctly.  

\section{HEAVY QUARK HADROPRODUCTION} 
\def\GeV{{\rm GeV}}

\def\figHQdata{
\begin{figure}[htbp]
\begin{center}
\leavevmode
 \epsfxsize=3in  \epsfbox{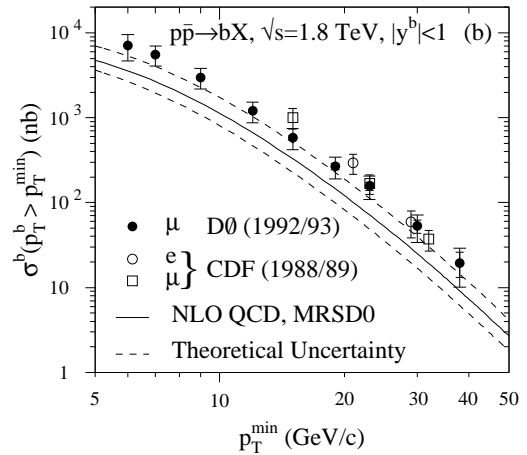}\\
 \epsfxsize=3in \epsfbox{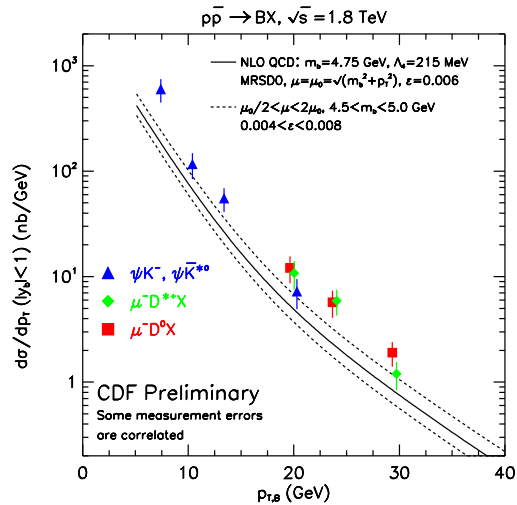}
\end{center}
      \caption{
Heavy quark hadroproduction data. 
{\it Cf.}, Ref.~\protect\cite{hqdata}.
}
   \label{fig:figHQdata}
\end{figure}
}

\def\figcacciari{
\begin{figure}[htbp]
\begin{center}
\leavevmode
 \epsfxsize=3in  \epsfbox{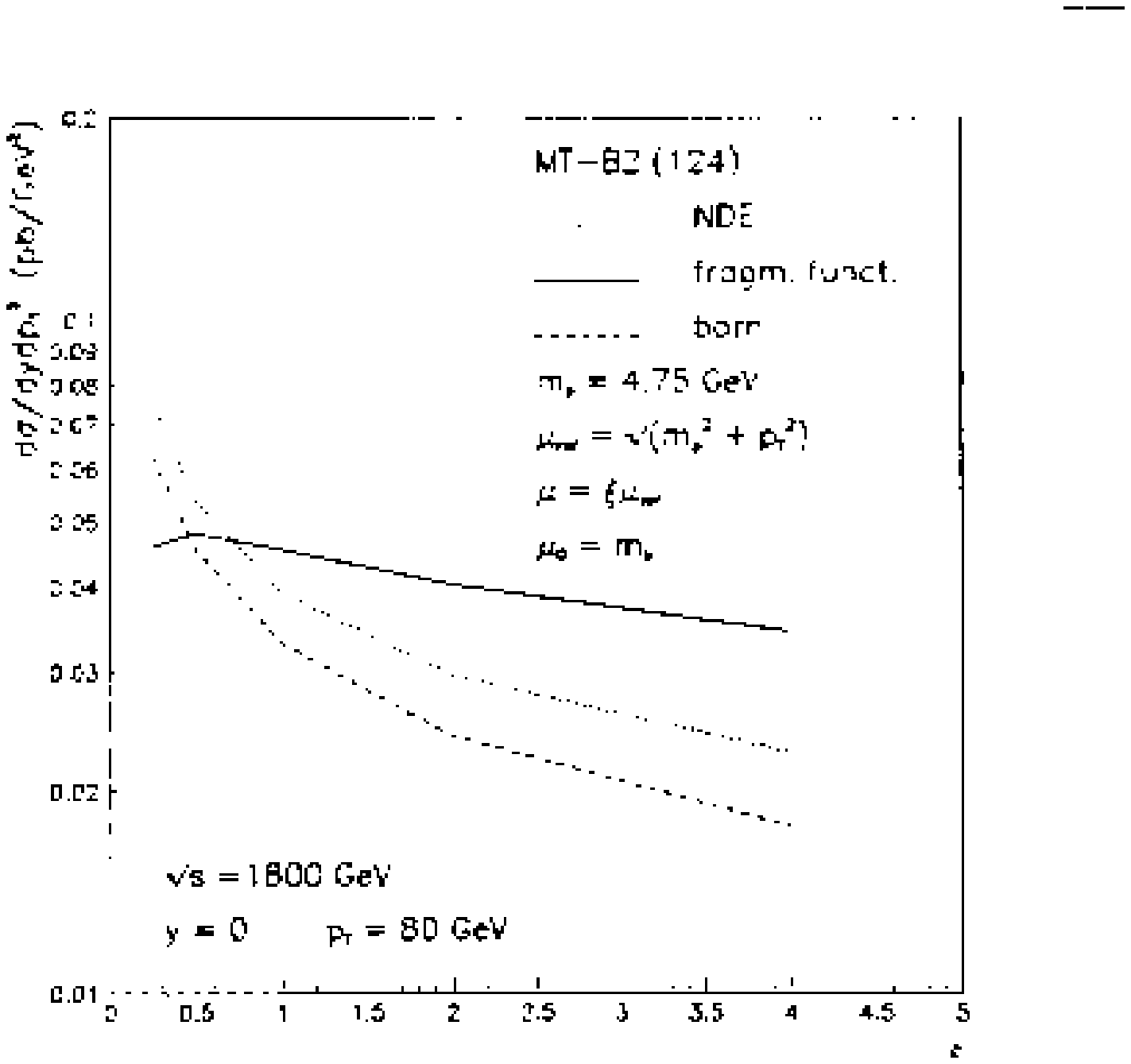}
\end{center}
      \caption{
Scale dependence of the heavy quark hadroproduction cross section
as a function of  $\mu = \xi \mu_{ref}$ at $y=0$ and $p_t= 80\, \GeV$. 
The NDE curve is the calculation of 
Ref.~\protect\cite{nde}.
The {\it fragm., funct.} and {\it born} curves are the calculation of
Ref.~\protect\cite{Greco}. }
   \label{fig:figcacciari}
\end{figure}
}

\def\figProd{
\begin{figure}[htbp]
\begin{center}
\leavevmode
\hbox{
 \epsfxsize=0.45\textwidth  \epsfbox{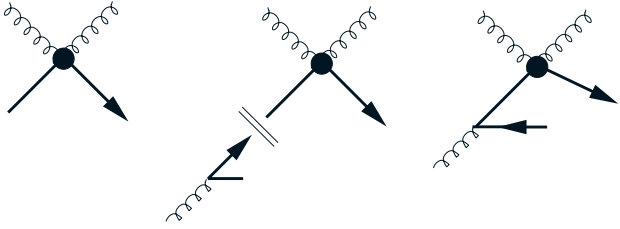}
}
\end{center}
      \caption{
a)~Generic leading-order diagram for flavor-excitation (LO-FE), $gQ\to gQ$.
b)~Subtraction diagram for flavor-excitation (SUB-FE),
   ${}^1f_{g\to Q} \otimes \sigma(gQ\to gQ)$.
c)~Next-to-leading-order diagram for flavor-creation (NLO-FC).
\null\hfill\null}
   \label{fig:figProd}
\end{figure}
}

\def\figDecay{
\begin{figure}[htbp]
\begin{center}
\leavevmode
\hbox{
 \epsfxsize=0.45\textwidth  \epsfbox{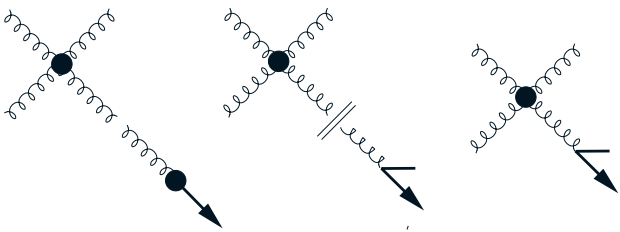}
}
\end{center}
      \caption{
a)~Generic leading-order diagram for flavor-fragmentation (LO-FF),
   $\sigma(gg\to gg) \otimes D_{g\to Q}$.
b)~Subtraction diagram for flavor-fragmentation (SUB-FF),
   $\sigma(gg\to gg) \otimes {}^1d_{g\to Q}$.
c)~Next-to-leading-order diagram for flavor-creation (NLO-FC).
\null\hfill\null}
   \label{fig:figDecay}
\end{figure}
}
\def\figFeSub{
\begin{figure}[htbp]
\begin{center}
\leavevmode
 \hbox{
 \epsfxsize=0.20\textwidth  \epsfbox{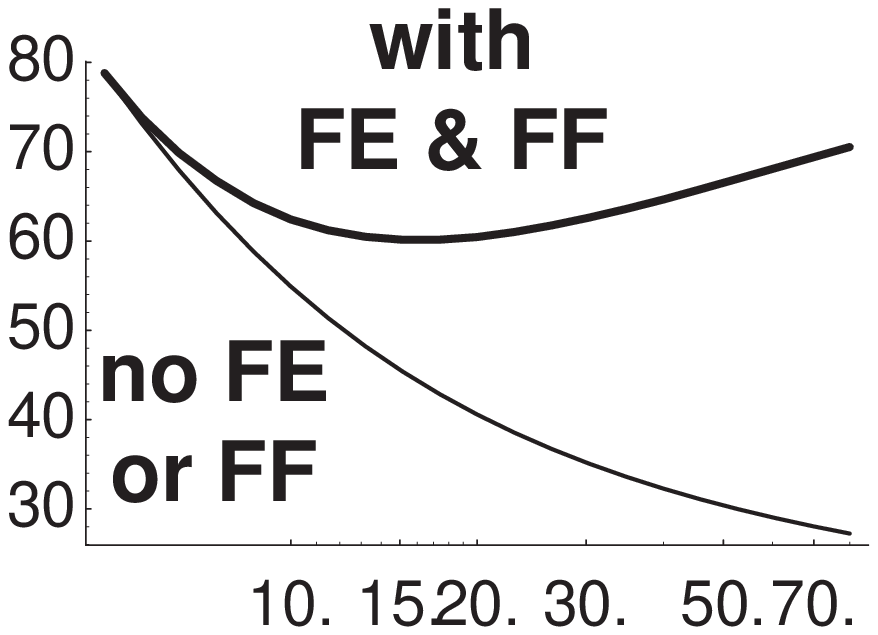}
 \hfill 
 \epsfxsize=0.25\textwidth  \epsfbox{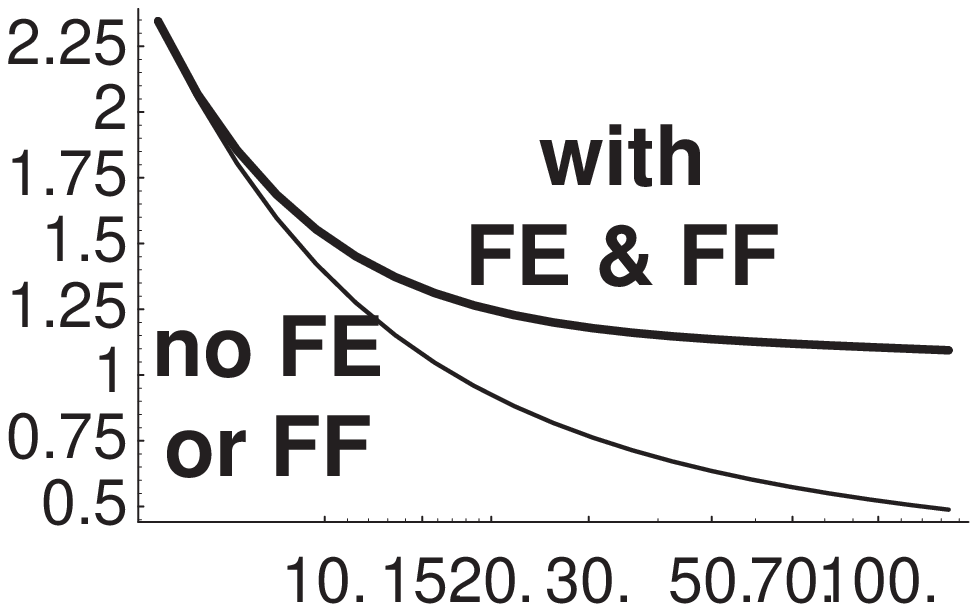}
}
\end{center}
      \caption{
The differential cross section $d^2 \sigma/dp_T^2/dy_1$ at
$p_T=20, \, 40 \, GeV$ and $y_1=0$ in  $(pb/GeV^2)$ {\it vs.} $\mu$.
 The lower curves (thin line) are the heavy quark
production cross sections {\it ignoring}
flavor-excitation (FE) and flavor-fragmentation (FF).
 The upper curves (thick line) are the heavy quark
production cross sections {\it including}
FE  and FF. {\it Cf.}, Ref.~\protect\cite{cost}.
\null\hfill\null}
   \label{fig:figfesub}
\end{figure}
}

 \figHQdata
 \figcacciari
 \figProd 
 \figDecay 
 \figFeSub

Improved experimental measurements of heavy quark hadroproduction have
increased the demand on the theoretical community for more precise
predictions.\cite{hqdata} The first Next-to-Leading-Order (NLO)
calculations of charm and bottom hadroproduction cross sections were
performed some years ago.\cite{nde} As the accuracy of the data
increased, the theoretical predictions displayed some shortcomings:
 1) the theoretical cross-sections fell well short of the measured values,  
 and 
 2) they displayed a strong dependence on the unphysical renormalization
scale  $\mu$. 
Both these difficulties indicated that these predictions were missing
important physics.

One possible solution for these deficiencies was to consider
contributions from large logarithms associated with the new quark mass
scale, such as\footnote{Here, $m_Q$ is the heavy quark mass, $s$ is
the energy squared, and $p_T$ is the transverse momentum.}
$\ln(s/m_Q^2)$ and $\ln(p_T^2/m_Q^2)$, Pushing the calculation to one
more order, formidable as it is, would not improve the situation since
these large logarithms persist to every order of perturbation theory.
Therefore, a new approach was required to include these logs.
 
In 1994, Cacciari and Greco\cite{Greco} observed that since the heavy
quark mass played a limited dynamical role in the high $p_t$ region,
one could instead use the massless NLO jet calculation convoluted with
a fragmentation into a massive heavy quark pair to more accurately
compute the production cross section in the region $p_t \gg m_Q$. In
particular, they find that the dependence on the renormalization scale
is significantly reduced, (cf., Fig.~\ref{fig:figcacciari}).

A recent study\cite{cost} investigated using initial-state heavy quark
PDFs and final-state fragmentation functions to resum the large
logarithms of the quark mass.  The principle ingredient was to include
the leading-order flavor-excitation (LO-FE) graph (Fig.~\ref{fig:figProd})
and the leading-order flavor-fragmentation (LO-FF) graph
(Fig.~\ref{fig:figDecay}) in the traditional NLO heavy quark
calculation.\cite{nde} These contributions can not be added naively to
the ${\cal O}(\alpha_s^3)$ calculation as they would double-count
contributions already included in the NLO terms; therefore, a
subtraction term must be included to eliminate the region of phase
space where these two contributions overlap.  This subtraction term
plays the dual role of eliminating the large unphysical collinear logs
in the high energy region, and minimizing the renormalization scale
dependence in the threshold region.  The complete calculation
including the contribution of the heavy quark PDFs and fragmentation
functions 1) increases the theoretical prediction, thus moving it
closer to the experimental data, and 2) reduces the $\mu$-dependence
of the full calculation, thus improving the predictive power of the
theory. (Cf., Fig~\ref{fig:figfesub}.)

In summary, heavy quark hadroproduction is of interest experimentally
because of the wealth of data allows precise tests of many different
aspects of the theory, namely radiative corrections, resummation of
logs, and multi-scale problems.  Hence, this is a natural testing
ground for QCD, and will allow us to extend the region of validity for
the heavy quark calculation.  This is an essential step necessary to
bring theory in agreement with experiment.

\section{SUMMARY} 
\subsection{Kinematic Reach of Future Machines} \label{sec:reach}

\def\figmapi{
\begin{figure}[t]
\begin{center}
 \leavevmode
  \epsfxsize=3in  
  \epsfbox{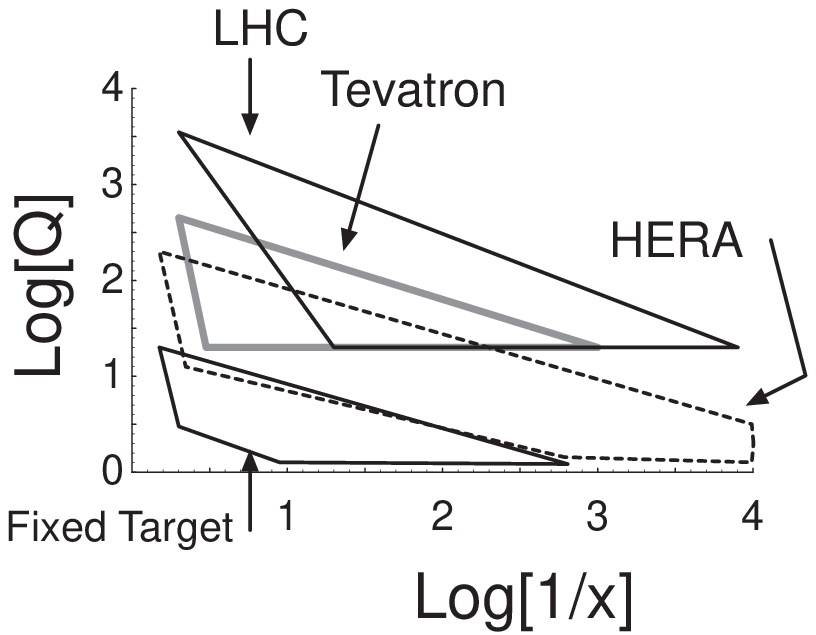}
 \end{center}
      \caption{Kinematic reach of present and planned facilities.  
               Note the full $\{x,Q^2\}$ region is clipped by the plot.
 }
   \label{fig:figmapi}
\end{figure}
}
\def\figmapii{
\begin{figure}[t]
\begin{center}
 \leavevmode
  \epsfxsize=3in  
  \epsfbox{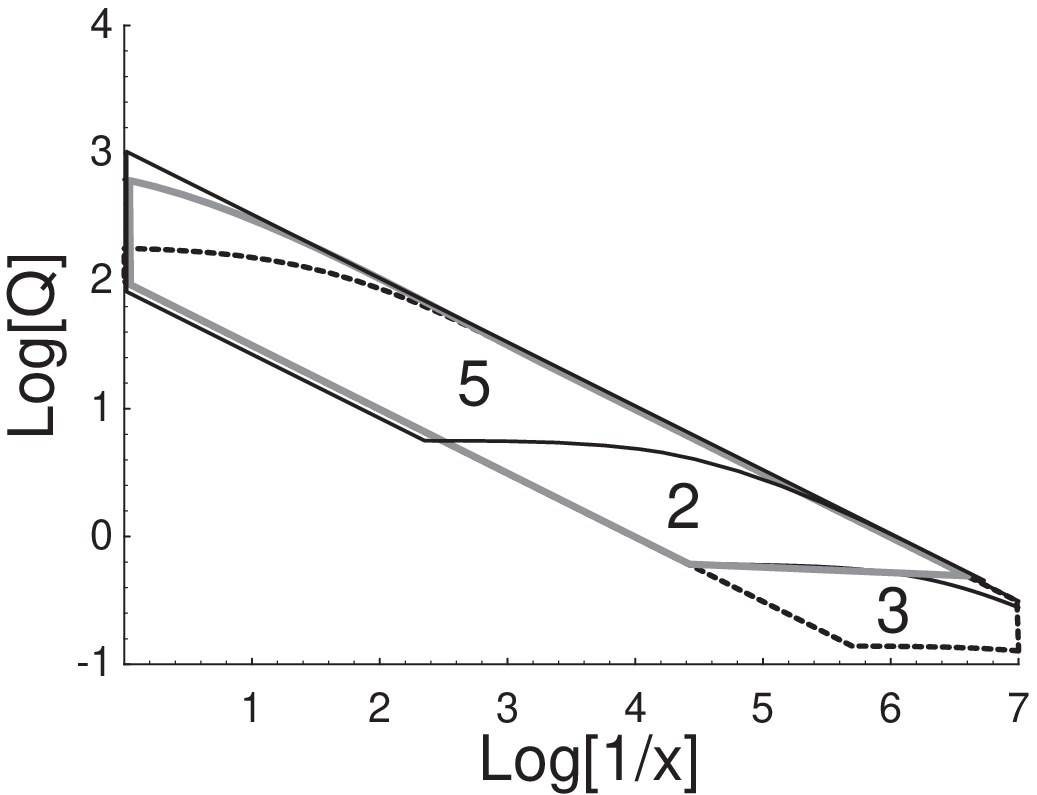}
 \end{center}
       \caption{Kinematic reach of future facilities.
 }
   \label{fig:figmapii}
\end{figure}
}
\def\figmapiii{
\begin{figure}[t]
\begin{center}
 \leavevmode
  \epsfxsize=3in  
  \epsfbox{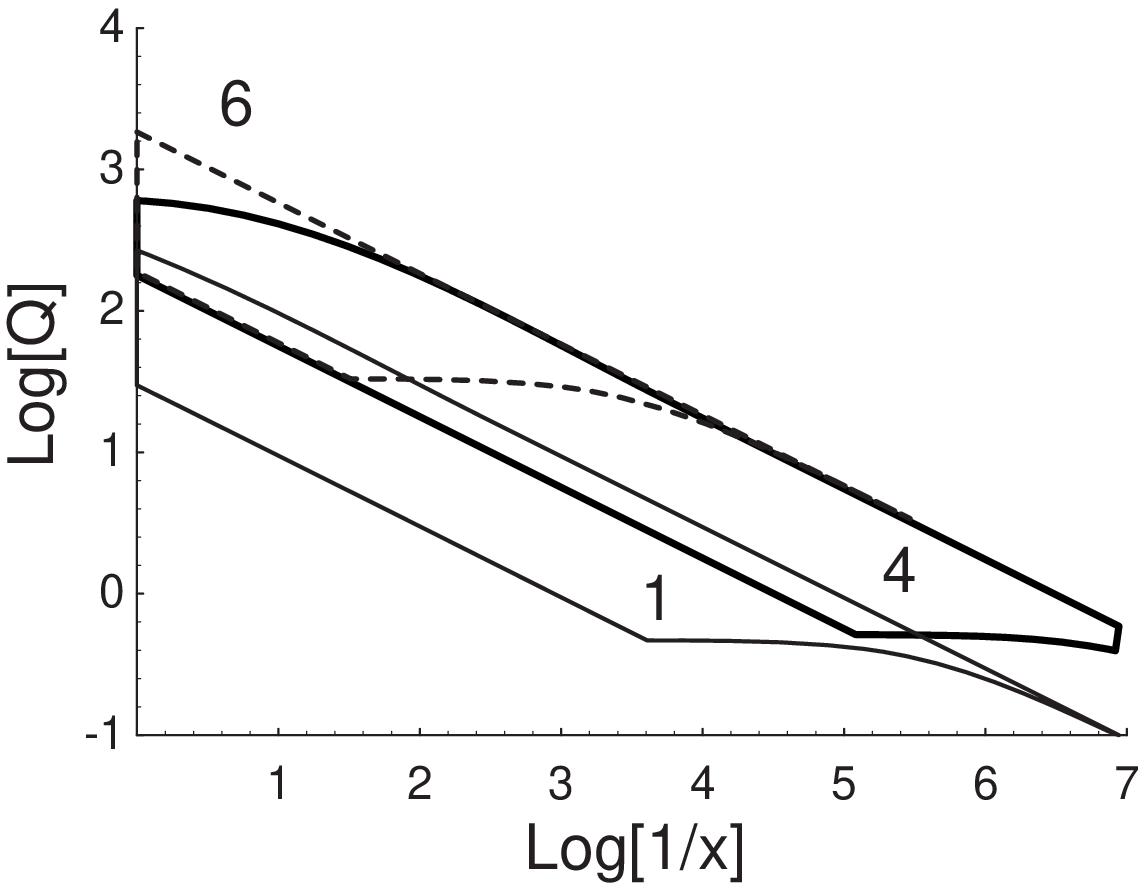}
 \end{center}
       \caption{Kinematic reach of future facilities.
 }
   \label{fig:figmapiii}
\end{figure}
}
\def\tablemap{
\begin{table}[htbp]
\begin{center}
\small
\caption{
Future $ep$ collider machines chosen for study. 
}

\vskip 10pt

\label{tablemap}
\begin{tabular}{|c|r|r|r|c|}
\hline
Index & E$_{\rm lepton}$ & E$_{\rm proton}$ & $\sqrt{s}$
& Machine(s)\\ 
 & (Gev) & (Gev) & (Gev) & \\
\hline\hline
1 &   27 &    820 &  300 & Hera\\ \hline
2 &   35 &  7,000 &  990 & Lep $\times$ LHC\\ \hline
3 &    8 & 30,000 &  980 & Low E lepton $\times$ 60 GeV pp\\ \hline
4 &   30 & 30,000 & 1900 & Lep $\times$ 60 GeV pp\\ \hline
5 &  500 &    500 & 1000 & NLC $\times$ conv. p\\ \hline
6 & 2,000 &   500 & 2000 & $\mu$ collider $\times$ conv. p\\
\hline
\end{tabular}

\end{center}
\normalsize
\end{table}
}

 \tablemap
 \figmapii
 \figmapiii
 \figmapi

A central goal of this workshop was to study the physics 
potential of future facilities.  Here, we focus on 
lepton-hadron colliders. We expand our study beyond the single $ep$ machine
proposed in the workshop outline, and consider a mix of lepton and 
hadron beams from those proposed for the lepton-lepton and hadron-hadron options. 
The complete list   is given in Table~\ref{tablemap}. 
 To covert these parameters into the $\{x,Q^2\}$ range, we make use of:
\begin{equation}
y = 1 - \frac{E'_{e}}{2E_{e}}(1 - \cos\theta_{\ell})
\quad ,
\end{equation}
\begin{equation}
Q^2 = 2 E_{e} E'_{e} (1 + \cos\theta_{\ell})
\quad ,
\end{equation}
and
\begin{equation}
x = Q^2/sy
\quad .
\end{equation}
For collider kinematics, we use
\begin{equation}
s \sim 4 E_{e} E_{p}
\quad .
\end{equation}
Here,  
$E_{e}$ is the incoming lepton energy, 
$E'_{e}$ is the outgoing lepton energy, 
$E_{p}$ is the incoming hadron energy, 
and $\theta_{\ell}$ is the lepton scattering angle.

To set practical limits on measurement of the final state, we impose: 
\begin{itemize}
\item $y > 0.01$  \quad (resolution), 
\item $y<1$  \quad (kinematic cut)
\item $\theta_{\ell} > 10^{\circ}$
\item $\theta_{\ell} < 179^{\circ}$
\end{itemize}
The constraint  $\theta_{\ell} < 179^{\circ}$ may be somewhat optimistic; 
if we relax this to  $\theta_{\ell}  \lsim 176^{\circ}$, the result
is to lose some of the low $Q$ region.  
 The constraint  $\theta_{\ell} > 10^{\circ}$ has a relatively small effect;
for the higher energy machines ({\it e.g}, 2 \& 3), it clips the upper $Q$ region.

We display the kinematic reach for these proposed machines in 
Figs.~\ref{fig:figmapii} and \ref{fig:figmapiii}.  We include HERA for reference. 
In Fig.~\ref{fig:figmapii}, we show the three machine options with a CMS of 
$\sqrt{s} \sim 1\, {\rm TeV}$.  
In Fig.~\ref{fig:figmapiii}, we show HERA and the remaining two  machine options.
In Fig.~\ref{fig:figmapi}, we show the present and planned (LHC) facilities. 

     Although there is currently no plan to extract the primary beam to 
make a neutrino fixed target experiment at either the LHC or at a 2 TeV muon 
collider, there is a case to be made for doing precisely that.   
First of all, it would be very interesting to see if there were 
an anomalous rise in $xF_3$ similar to that seen in $F_2$ at HERA.  
Secondly, the low $x$ region of the Bjorken integral is anomalously 
large, and an outstanding question is, what is the very low $x$ behavior of 
the Gross-Llewellyn Smith integral ($xF_3$)?  Either an experiment at 
the LHC or one at a 2 TeV muon collider could extend the 
range of the "Fixed Target" region indicated in Figure 23 by an order of 
magnitude in the log ($1/x$) direction, assuming an order of magnitude 
higher neutrino energies than what CCFR/NuTeV has.  The neutrino cross 
section would be an order of magnitude higher than the one applicable for 
CCFR/NuTeV, so good statistics are in principle attainable.  Although 
these experiments would not have the kinematic reach to extremely low 
$x$ that $ep$ machines have, they can measure to high precision the 
non-singlet structure function, which at present has only been measured 
down to $x=.01$.  In principle an ep machine running with both positive and 
negative leptons could do the same, but the luminosity requirements may 
be prohibitively high.  We have still not learned all that we can learn 
from neutrino experiments, and even modest improvements in neutrino 
energies can uncover much new ground.  

While we would of course like to probe the full $\{x,Q^2\}$ space, there
are some particular reasons why the small $x$ region is of special
interest.  For example, the rapid rise of the $F_2$ structure function
observed at HERA suggests that we may reach the parton density
saturation region more quickly than anticipated.  Additionally, the
small $x$ region can serve as a useful testing ground for BFKL,
diffractive phenomena, and similar processes.  We can clearly see in
Fig.~\ref{fig:figmapii} that with a fixed $\sqrt{s}$, we can best
probe the small $x$ region with a high energy hadron beam colliding
with a low energy lepton beam, and the loss in the high $Q$ region is
minimal.  From these (preliminary) studies, it would seem the optimal
$ep$ facility would match the highest energy hadron beam
available with a modest energy lepton beam.


\end{document}